\documentclass[amsmath,amssymb,twocolumn]{revtex4-1}
\usepackage{txfonts, color}
\usepackage{amsmath, amssymb, bm,physics}

\usepackage{graphicx}
\newcommand{\us}{\uparrow}
\newcommand{\ds}{\downarrow}
\usepackage{hyperref}

\begin{document}
\title{Theory of Magnetic-Texture-Induced Anomalous Hall Effect on the Surface of Topological Insulators}
\author{Terufumi Yamaguchi}
\author{Ai Yamakage}
\affiliation{$^1$Yukawa Institute for Theoretical Physics, Kyoto University, Kyoto 606-8502, Japan\\
$^2$Department of Physics, Nagoya University, Nagoya 464-8602, Japan
}

\begin{abstract}
    The anomalous Hall effect is caused by magnetic textures such as skyrmions. 
	We derive an analytical formula of the Hall conductivity on the surface of a topological insulator 
	up to third order in magnetization, $\boldsymbol{M}(\boldsymbol{x})$, based on a perturbative approach.
	We identify the magnetic textures that contribute to the Hall conductivity up to third order in magnetization and second order in spatial differentiation.
	We treat magnetization as a perturbation 
	to calculate the Hall conductivity for each magnetic texture based on the linear response theory.
Furthermore, we estimate the skyrmion-induced Hall conductivity and confirm that it depends on the shape of skyrmions, such as Bloch-type or N\'eel-type skyrmions. The results of this study can be applied not only to conventional skyrmion systems but also to more general magnetic structures.
\end{abstract}

\maketitle


\textit{Introduction}.--
The Hall effect without a magnetic field is known as the anomalous Hall effect (AHE) \cite{Sinitsyn2008, Nagaosa2010a, Chang2013a, Qi2020}. 
It occurs in a wide variety of materials, and it is desired for applications in spintronic devices \cite{Kondou2016a, Rojas-Sanchez2016}.
The AHE has been mainly investigated in spin--orbit coupled systems, 
in which an emergent magnetic field originates from a topological structure in a momentum space.
In recent years, the AHE has been experimentally and theoretically examined, 
particularly in antiferromagnets such as Mn$_{3}$Sn\cite{Nakatsuji2015} and Mn$_{3}$Ge \cite{Kanazawa2011a, Kiyohara2016, Nayak2016a},
becoming a central subject of antiferromagnetic spintronics.

The AHE caused by a spatial variation in magnetization and not spin--orbit coupling is referred to as the topological Hall effect (THE) \cite{Ye1999, Bruno2004, Neubauer2009, Nakazawa2018, Ishizuka2018}. 
The THE is induced by an emergent magnetic field that originates from a topological structure in a real space.
In ferromagnetic metals without spin--orbit coupling, the THE is proportional to the skyrmion number, $(1/4 \pi)\int {\rm d} \bm{r} \bm{M} \cdot (\partial_{x} \bm{M} \times \partial_{y} \bm{M})$, where $\bm{M}$ is the unit vector that points in the direction of the localized spin.
The AHE and THE have been studied in the presence and absence of spin--orbit coupling, respectively. 
However, few studies have investigated the Hall effect caused by the spatial variation in magnetization  in the presence of spin--orbit coupling.
There are experimental \cite{yasuda16} and pioneering theoretical \cite{araki17, Wang2020} studies on the Hall effect induced by the magnetic textures on the surface of topological insulators, which are representative systems that show extremely strong spin--orbit interactions.
These studies assume only magnetic skyrmions, and it is not clear which magnetic texture causes the Hall effect.

In this study, we investigate the AHE due to an arbitrary spatial variation in magnetization on the surface of a topological insulator. 
First, from the viewpoint of symmetry, we determine the magnetic textures that induce the AHE and whether the Hall conductivity is an even or odd function of the chemical potential. 
Next, we calculate the Hall conductivity based on the linear response theory. 
We treat the coupling between electrons and magnetization as a perturbation and calculate the AHE up to third order in magnetization. 
Finally, we show that the AHE in a topological insulator depends on the type of skyrmions, i.e., Bloch-type or N\'{e}el-type skyrmions, whereas that in conventional ferromagnets without spin--orbit coupling does not.


\textit{Microscopic model}.--
A microscopic model consists of three parts: conduction electrons on the surface of a topological insulator, $H_{0}$, 
nonmagnetic impurities, $H_{\rm imp}$, and the coupling between the conduction electrons and magnetization, $H_{sd}$ \cite{sakai14}.
\begin{align}
 H_0 &= \sum_{\boldsymbol{k}} c^\dag_{\boldsymbol{k}} \qty[v_{\mathrm{F}} \qty(k_y s_x - k_x s_y) - \mu] c_{\boldsymbol{k}}, 
 \\
 H_{\mathrm{imp}}
 &= \int d\boldsymbol{r} c^\dag(\boldsymbol{r}) V_{\mathrm{imp}}(\boldsymbol{r}) c(\boldsymbol{r}),
 \\
 H_{sd} &= \int d\boldsymbol{r} c^\dag(\boldsymbol{r}) 
 \qty[-J_{z} M_z s_z - J_{\perp} \qty(M_x s_x + M_y s_y)] c(\boldsymbol{r}),
\end{align}
where $c^{\dagger}_{\bm{k}} = (c^{\dagger}_{\bm{k},\us} , c^{\dagger}_{\bm{k},\ds})$ 
and $c^\dag(\boldsymbol{r}) = S^{-1/2} \sum_{\boldsymbol{k}} e^{- i \boldsymbol{k} \cdot \boldsymbol{r}} c^\dag_{\boldsymbol{k}}$ 
are the spinors of the creation operator in the momentum ($\bm{k} = (k_{x}, k_{y})$) and real spaces, respectively, 
$v_{\rm F}$ is the Fermi velocity, $s^{\alpha} \ (\alpha = x,y,z)$ denotes the Pauli matrices, $\mu$ is the chemical potential, 
$V_{\rm imp}(\bm{r}) = \sum_{j} u_{\rm i} \delta (\bm{r} - \bm{R}_{j})$ is the impurity potential, 
and $J_{z}$ and $J_{\perp}$ are the \textit{s-d} exchange coupling constants with the $z$ ($x,y$) component of magnetization, $M_{z} \ (M_{x}, M_{y})$.


\textit{Symmetry consideration}.-- 
The minimal model with a linear dispersion has $C_{\infty v}$ point-group symmetry, i.e., continuous rotational symmetry along the $z$ axis and vertical mirror-reflection symmetry. 
The conductivity tensor, $\sigma_{ij}$, is decomposed into the irreducible representations (irreps) of $C_{\infty v}$ as $A_1: \sigma_{xx}+\sigma_{yy}$, $A_2: \sigma_{xy}-\sigma_{yx}$, and $E_2: (\sigma_{xx}-\sigma_{yy}, \sigma_{xy}+\sigma_{yx})$.
These conductivities can be induced by an external field that belongs to the same irrep, as listed in Table \ref{irrep}. 

The linear model shows ``gauge'' symmetry.
The in-plane magnetic texture ($M_x$ and $M_y$) is equivalent to the ``gauge'' field, $\tilde{\boldsymbol{A}} = (ev_{\mathrm{F}})^{-1} J_\perp (M_y, -M_x)$, which is minimally coupled to the system. 
This implies that physical quantities are the functions of only ``magnetic'' field $\tilde B_z = \partial_x \tilde A_y - \partial_y \tilde A_x \propto \boldsymbol{\nabla}\cdot \boldsymbol{M}$ and $M_z$ owing to ``gauge'' invariance \cite{sakai14}.
Table \ref{irrep} shows the gauge-invariant irreps of magnetic textures. 
\begin{table*}
 \caption{Gauge-invariant functions of magnetic textures. $f(\Delta)$ is an odd function of $\Delta$, which belongs to the $A_2$ irreducible representation (Irrep). }
 \begin{tabular}{llllllll}
 	\hline
 	Irrep & $\sigma_{ij}$ & $\boldsymbol{\nabla}\boldsymbol{M}$ & $\boldsymbol{\nabla}\boldsymbol{\nabla}\boldsymbol{M}$ & $\boldsymbol{M}\boldsymbol{\nabla}\boldsymbol{M}$ 
 	& $\boldsymbol{\nabla}\boldsymbol{M}\boldsymbol{\nabla}\boldsymbol{M}$ 
 	& $\boldsymbol{M} \boldsymbol{M} \boldsymbol{\nabla} \boldsymbol{M}$
 	& $\boldsymbol{M} \boldsymbol{\nabla} \boldsymbol{M} \boldsymbol{\nabla} \boldsymbol{M}$
 	\\
 	\hline
 	$A_2$ & $\sigma_{xy}-\sigma_{yx}$ & $\boldsymbol{\nabla} \cdot \boldsymbol{M}$ & $\nabla^2 M_z$ & $f(\Delta) M_z \boldsymbol{\nabla} \cdot \boldsymbol{M}$, 
 	& 
 	$f(\Delta) (\boldsymbol{\nabla} M_z)^2$, $f(\Delta) (\boldsymbol{\nabla} \cdot \boldsymbol{M})^2$, 
 	& $M_z^2 \boldsymbol{\nabla} \cdot \boldsymbol{M}$
 	& $M_z\qty(\boldsymbol{\nabla}M_z)^{2}$,  $M_z\qty(\boldsymbol{\nabla} \cdot \boldsymbol{M})^2$
 	\\
 	\hline
 \end{tabular}
 \label{irrep}
\end{table*} 

The minimal model also has particle-hole symmetry, $s_x H^*(\mu, M_x, M_y, M_z) s_x = -H(-\mu, -M_x, -M_y, M_z)$, whereas electric current $\boldsymbol{j}$ does not, $s_x \boldsymbol{j}^* s_x = \boldsymbol{j}$. 
On the basis of these relations, we obtain
\begin{align}
	&
 \sigma_{ij}(\mu, T, M_x, M_y, M_z) 
 = 
 \sigma_{ij}(-\mu, -T, -M_x, -M_y, M_z).
\end{align}
A detailed derivation of the above relation is given in Appendix A.
We have verified that all these symmetries hold in the calculated results shown below.


\textit{Microscopic calculation}.--
To calculate the Hall conductivity, we write the nonperturbative retarded Green's function in this model as
(assuming point-like impurities, $V_{\rm imp} = u_{0} \sum_{i} \delta (\bm{r} - \bm{r}_{i})$, 
and applying the Born approximation to self-energy)
\begin{align}
	G^{R}_{\bm{k}}(\varepsilon)
	=
	g^{0}_{\bm{k}}(\varepsilon) + \sum_{\alpha = x,y,z} g^{\alpha}_{\bm{k}}(\varepsilon) s^{\alpha},
	\label{eq:Green}
\end{align}
where $g^{0}_{\bm{k}}(\varepsilon) = (\varepsilon + \mu + i \gamma)/D_{\bm{k}}(\varepsilon)$,
$g^{x}_{\bm{k}}(\varepsilon) = - v_{\rm F} k_{y}/D_{\bm{k}}(\varepsilon)$,
$g^{y}_{\bm{k}}(\varepsilon) = v_{\rm F} k_{x}/D_{\bm{k}}(\varepsilon)$,
$g^{z}_{\bm{k}}(\varepsilon) = - (\Delta - i (\Delta/\mu)\gamma)/D_{\bm{k}}(\varepsilon)$,
$D_{\bm{k}}(\varepsilon) 
= (\varepsilon + \mu + i \gamma)^{2} - (\Delta - i (\Delta/\mu) \gamma)^{2} - v_{\rm F}^{2} k^{2}$,
$\gamma = \gamma_{0} |\mu| \Theta (|\mu| - |\Delta|)$, and
$\gamma_{0} = n_{\rm i} u_{0}^{2} / (2 v_{\rm F})^{2}$.
Here, we introduce the gap, $\Delta$, as a contribution from an external magnetic field along the $z$ direction and the uniform $z$-component magnetization, $\int d\boldsymbol{r} M_z$.

\begin{figure}
	\centering
	\includegraphics[width=8.5cm]{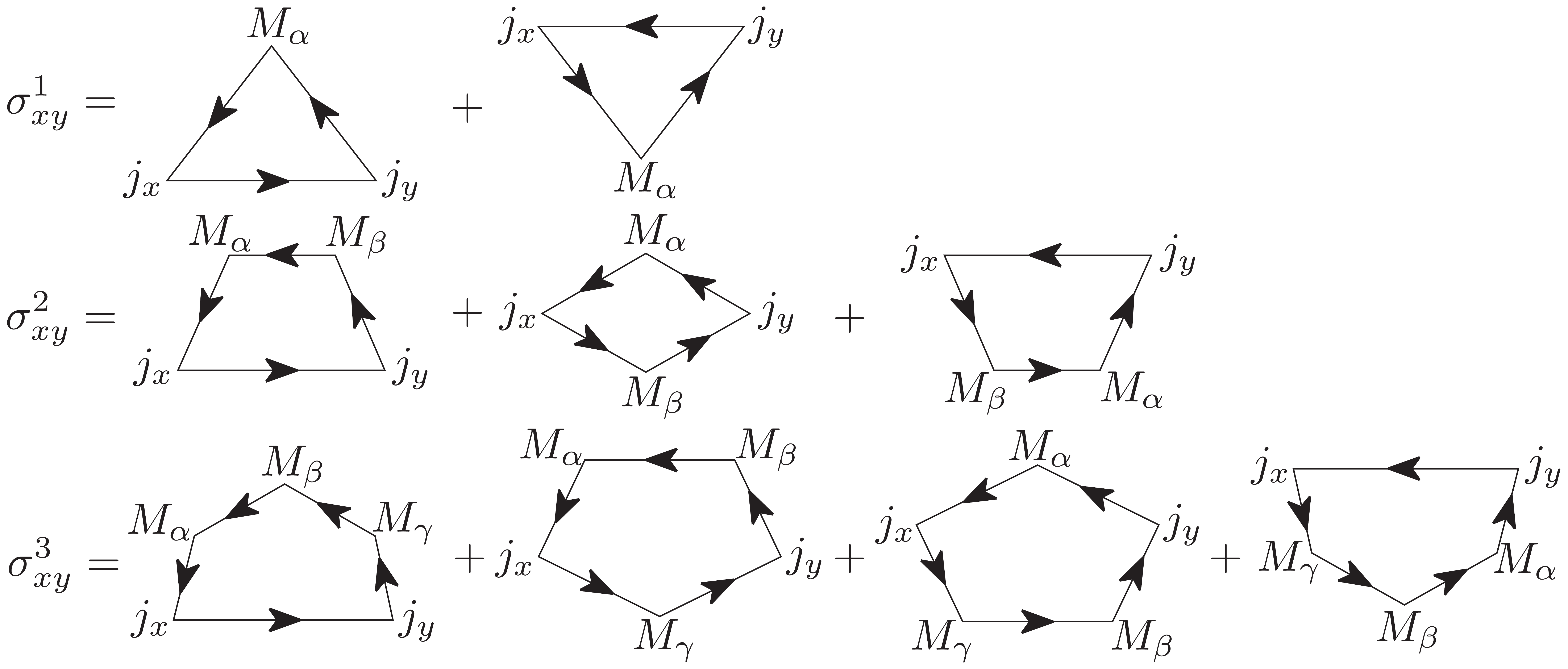}
	\caption{Feynman diagrams corresponding to $\sigma^{1}_{xy}, \sigma^{2}_{xy}$, and $\sigma^{3}_{xy}$.}
	\label{fig:diagram}
\end{figure}

We calculate the Hall conductivities induced by the magnetization up to the third order.
According to the linear response theory, 
the Hall conductivity for $n(=1,2,3)$-th order in magnetization, $\sigma^{n}_{\rm Hall}$, can be expressed as
\begin{align}
	\sigma^{n}_{\rm Hall}
	=&
	\left( \sigma^{n}_{xy} - \sigma^{n}_{yx} \right)/2,
	\label{eq:sigma_Hall}
	\\
	\sigma^{n}_{\alpha \beta}
	=&
	\lim_{\omega \to 0} \frac{K^{n}_{\alpha \beta}(\omega + i0) - K^{n}_{\alpha \beta}(0)}{i \omega},
	\label{eq:sigma_xy}
\end{align}
where $\omega$ is the frequency and we take the DC limit $\omega \to 0$. 
$K^{n}_{\alpha \beta}$ is defined as (with Matsubara frequency $\omega_{\lambda}$)
\begin{align}
	K^{n}_{\alpha \beta} (i \omega_{\lambda})
	=
	\int_{0}^{1/T} {\rm d} \tau 
	{\rm e}^{i \omega_{\lambda} \tau}
	\langle T_{\tau} j_{\alpha} (\tau) j_{\beta} \rangle_{n},
	\label{eq:K_xy}
\end{align}
where $T$ is the temperature,
$j_{\alpha} = v_{\rm F} (\hat{z} \times \bm{s})_{\alpha} (\alpha = x,y)$ is the current operator for the $\alpha$ component, 
$T_{\tau}$ is the (imaginary) time ordering,
and $\langle \cdots \rangle_{n}$ denotes the thermal (equilibrium) average with the $n$-th perturbation of magnetization.
The Feynman diagrams for $\sigma^{1}_{xy}$, $\sigma^{2}_{xy}$, and $\sigma^{3}_{xy}$ are shown in Fig.~\ref{fig:diagram}.
$\sigma^{1}_{xy} = \lim_{\Omega \to 0} \left( K^{1}_{xy}(\omega + i0) - K^{1}_{xy}(0) \right)/i \omega$ 
is expressed using Green's function defined in Eq.~(\ref{eq:Green}) as
\begin{align}
	K^{1}_{xy}(i \omega_{\lambda})
	=
	T \sum_{m,\bm{k}} \tr
	\left[ 
		j_{x} G^{+}_{+} s^{\alpha} G^{+} j_{y} G
		+ j_{x} G^{+} j_{y} G s^{\alpha} G_{-}
	\right] M_{\alpha}(\bm{Q}),
	\label{eq:K1}
\end{align}
where $G = G_{\bm{k}}(i \varepsilon_{m})$, $G^{+} = G_{\bm{k}}(i \varepsilon_{m} + i \omega_{\lambda})$,
$G^{+}_{+} = G_{\bm{k}+\bm{Q}}(i \varepsilon_{m} + i \omega_{\lambda})$, and $G_{-} = G_{\bm{k}-\bm{Q}}(i \varepsilon_{m})$. $\bm{Q}$ is the wavevector of magnetization, $\varepsilon_{m}$ is the fermionic Matsubara frequency, and $\alpha = x,y,z$ is the direction of magnetization.
Similarly, $K^{2}_{xy}$ and $K^{3}_{xy}$ are given by
\begin{align}
	K^{2}_{xy} (i \omega_{\lambda})
	=&
	T \sum_{m,\bm{k}}
	\tr
	\left[ 
		j_{x} G^{+} s^{\alpha} G^{+}_{+} s^{\beta} G^{+} j_{y} G
		+ j_{x} G^{+} s^{\alpha} G^{+}_{+} j_{y} G_{+} s^{\beta} G
	\right.
	\notag \\
	&
	\left.
	+ j_{x} G^{+} j_{y} G s^{\alpha} G_{+} s^{\beta} G
	\right]
	M_{\alpha} (\bm{Q}) M_{\beta} (-\bm{Q}),
	\label{eq:K2}
	\\
	K^{3}_{xy} (i \omega_{\lambda})
	=&
	T \sum_{m,\bm{k}}
	\tr
	\left[
		j_{x} G^{+} s^{\alpha} G^{+}_{+} s^{\beta} G^{+}_{+ +} s^{\gamma} G^{+} j_{y} G
	\right.
	\notag \\
	&
	\left.
		+ j_{x} G^{+} s^{\alpha} G^{+}_{+} s^{\beta} G^{+}_{+ +} j_{y} G_{+ +} s^{\gamma} G
	\right.
	\notag \\
	&
	\left.
		+ j_{x} G^{+} s^{\alpha} G^{+}_{+} j_{y} G^{+}_{+} s^{\beta} G_{+ +} s^{\gamma} G
	\right.
	\notag \\
	&
	\left.
		+ j_{x} G^{+} j_{y} G s^{\alpha} G_{+} s^{\beta} G_{+ +} s^{\gamma} G
	\right]
	\notag \\
	& \times
	M_{\alpha}(\bm{Q}) M_{\beta}(\bm{Q}') M_{\gamma} (-\bm{Q}-\bm{Q}'),
	\label{eq:K3}
\end{align}
where $G_{+ +} = G_{\bm{k}+\bm{Q} + \bm{Q}'}(i \varepsilon_{m})$ and 
$G^{+}_{+ +} = G_{\bm{k}+\bm{Q}+\bm{Q}'}(i \varepsilon_{m} + i \omega_{\lambda})$.
In these calculations, we assume that $|\bm{Q}|, |\bm{Q}'| \ll k_{\rm F}$ ($k_{\rm F}$ is the Fermi wavenumber of electrons), i.e., a smooth magnetic texture.


Here we provide a few remarks about the calculations before describing the results.
The conductivity calculated using the linear response theory is divided into the Fermi surface terms,
which are the main contributors in metallic states, 
and the Fermi sea terms, which are the main contributors in insulating states.
As we are interested in the Hall conductivity in a metallic state, 
we calculate the contributions from all terms for first order in magnetization 
and only the Fermi surface terms for second border and third order in magnetization.
It is subsequently shown that the contributions from the Fermi surface terms for second order and third order in magnetization 
strongly depend on the impurities, $\gamma_{0}$,
and these terms are the primary contributors in clean systems, i.e., $\gamma_{0} \ll 1$.
Noting that the contributions depending on impurities are essentially different from the Berry-phase contributions 
discussed in many systems\cite{Chen2014a,Zhang2020}, which do not depend on impurities, 
and the former contributions are often dominant in metallic states.

\begin{figure}
	\centering
	\includegraphics[width=8.5cm]{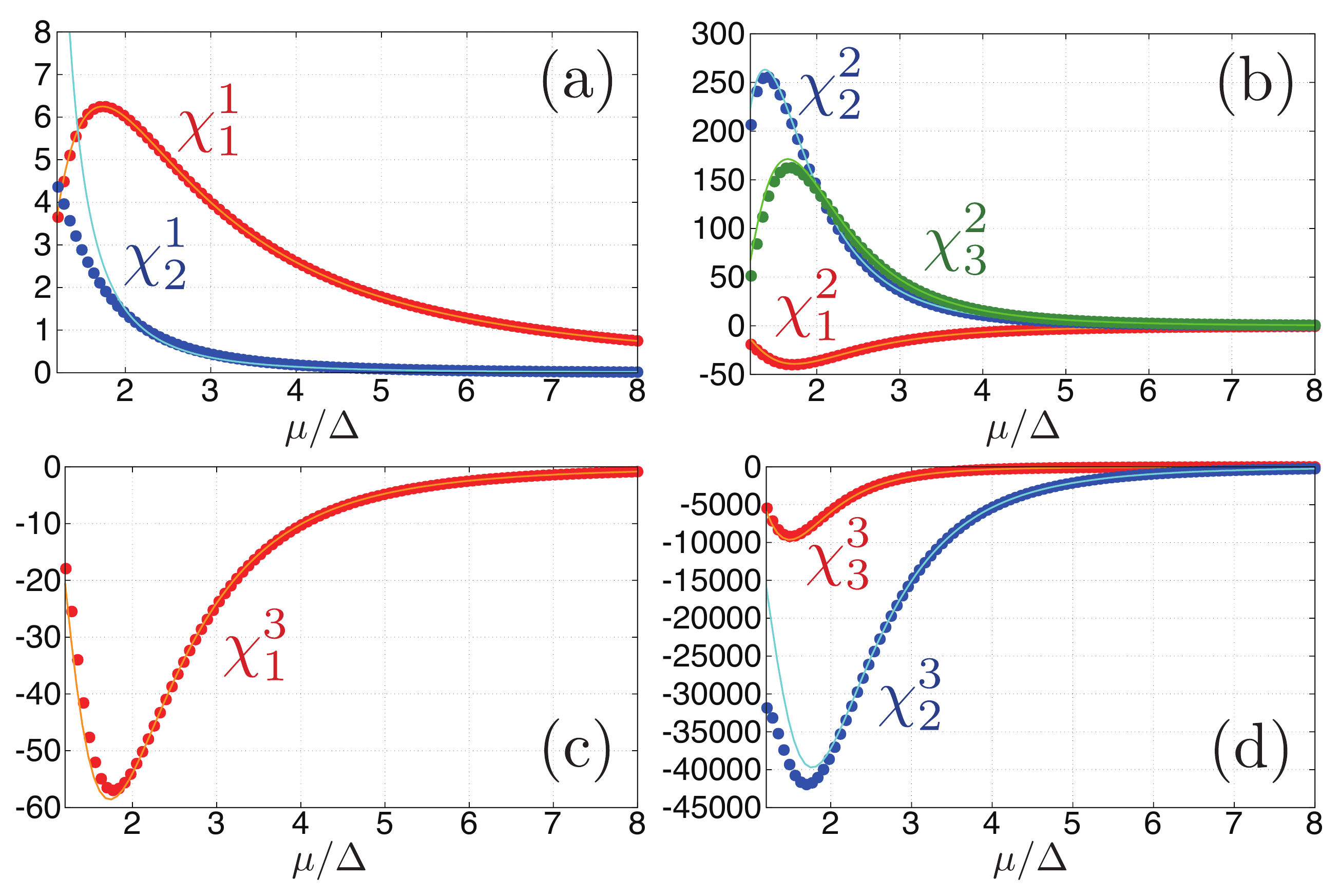}
	\caption{
        (Color online)
		Coefficients of Hall conductivity
		(a) $\chi^{1}_{1}$ and $\chi^{1}_{2}$, (b) $\chi^{2}_{1}, \chi^{2}_{2}$, and $\chi^{2}_{3}$,
		(c) $\chi^{3}_{1}$ and (d) $\chi^{3}_{2}$ and $\chi^{3}_{3}$, denoted by the dots.
		In this figure, we use $\gamma_{0} = 0.05$. 
		The solid lines represent the results obtained by neglecting higher-order $\gamma_{0}$.}
	\label{fig:Hallplot_all}
\end{figure}

The conductivity of the first order in magnetization, $\sigma^{1}_{\rm Hall}$, is divided into two parts, which are proportional to $\bm{\nabla} \cdot \bm{M}$ and $\nabla^{2} M_{z}$, as follows:
\begin{align}
	\sigma^{1}_{\rm Hall}
	=&
	\chi^{1}_{1} \frac{J_{\perp} v_{\rm F}}{|\Delta|^{2}}
	\int \frac{{\rm d} \bm{r}}{S} \bm{\nabla} \cdot \bm{M}
	+
	\chi^{1}_{2} \frac{J_{z} v_{\rm F}^{2}}{|\Delta|^{3}}
	\int \frac{{\rm d} \bm{r}}{S} \nabla^{2} M_{z},
	\label{eq:sigma_1}
\end{align}
where $\chi^{1}_{1}$ and $\chi^{1}_{2}$ are given by
\begin{align}
	\chi^{1}_{1}
	=&
	- \frac{1}{2 \pi}
	\Re
	\left[ 
		\zeta \eta \frac{i x + \gamma_{0} {\rm sgn}(\Delta) |x|}{\zeta^{2}}
	\right.
	\notag \\
	& \times
	\left.
		\left\{ 
			1 + 2 \eta \zeta + 4 \pi C_{\zeta} {\rm sgn}(\Delta) \eta |\eta| \zeta^{2}
		\right\}
	\right] - \frac{1}{96}
	\notag \\
	=&
	\frac{C_{\zeta} x (x^{2}-1)}{8 \gamma_{0}^{2} |x| (x^{2}+1)^{2}} 
	+ \mathcal{O} \left( \frac{1}{\gamma_{0}} \right),
	\label{eq:chi11}
	\\
	\chi^{1}_{2}
	=&
	- \frac{x^{2} - 1 + \gamma_{0}^{2} (x^{2} - 1) \Theta(|x|-1)}{\pi}
	\left[ 
		\frac{\eta \Re [\zeta]}{|\zeta|^{2}} + 2 \pi \eta |\eta| C_{\zeta}
	\right]
	\notag \\
	=&
	\frac{x^{2} (x^{2}+3)}{16 \pi \gamma_{0}^{2} |x|(x^{2}-1)(x^{2}+1)^{2}}
	+ \mathcal{O} \left( \frac{1}{\gamma_{0}} \right),
	\label{eq:chi12}
\end{align}
where $S$ is the system size (area), $x = \mu/\Delta$, 
$\zeta = \{ x + i \gamma_{0} |x| \Theta(|x|-1)) \}^{2} - \{1 - {\rm sgn}(\mu) i \gamma_{0} \Theta(|x|-1)\}^{2}$,
$\eta = 1/(2 \Im\zeta)$,
$C_{\zeta} = 1 - \chi_{\zeta}$ if $\Re \zeta > 0$, $C_{\zeta} = \chi_{\zeta}$ if $\Re \zeta < 0$,
and $\chi_{\zeta} = (1/\pi) \tan^{-1} (|\Im \zeta| / |\Re \zeta|)$.
$\chi^{1}_{1}$ and $\chi^{1}_{2}$ with $\gamma_{0} = 0.05$ are shown in Fig.~\ref{fig:Hallplot_all}(a)].
Note that $\chi^{1}_{1}$ and $\chi^{1}_{2}$ are even and odd functions of chemical potential $\mu$, respectively, because of particle-hole symmetry.


Similarly, $\sigma^{2}_{\rm Hall}$ and $\sigma^{3}_{\rm Hall}$ for $|\mu| > |\Delta|$ are written as
\begin{align}
	\sigma^{2}_{\rm Hall}
	=&
	\chi^{2}_{1} \frac{J_{\perp} J_{z} v_{\rm F}}{|\Delta|^{3}}
	\int \frac{\mathrm{d} \bm{r}}{S} M_{z} \bm{\nabla} \cdot \bm{M}
	+
	\chi^{2}_{2} \frac{J_{z}^{2} v_{\rm F}^{2}}{|\Delta|^{4}}
	\int \frac{\mathrm{d} \bm{r}}{S} \left( \bm{\nabla} M_{z} \right)^{2}
	\notag \\
	&+
	\chi^{2}_{3} \frac{J_{\perp}^{2} v_{\rm F}^{2}}{|\Delta|^{4}}
	\int \frac{\mathrm{d} \bm{r}}{S} \left( \bm{\nabla} \cdot \bm{M} \right)^{2},
	\label{eq:sigma_2}
	\\
	\sigma^{3}_{\rm Hall}
	=&
	\chi^{3}_{1} \frac{J_{\perp} J_{z}^{2} v_{\rm F}}{|\Delta|^{4}}
	\int \frac{\mathrm{d} \bm{r}}{S} M_{z}^{2} \bm{\nabla} \cdot \bm{M}
	+
	\chi^{3}_{2} \frac{J_{z}^{3} v_{\rm F}^{2}}{|\Delta|^{5}}
	\int \frac{\mathrm{d} \bm{r}}{S} M_{z} \left( \bm{\nabla} M_{z} \right)^{2}
	\notag \\ &
	+
	\chi^{3}_{3} \frac{J_{\perp}^{2} J_{z} v_{\rm F}^{2}}{|\Delta|^{5}}
	\int \frac{\mathrm{d} \bm{r}}{S} M_{z} \left( \bm{\nabla} \cdot \bm{M} \right)^{2},
	\label{eq:sigma_3}
\end{align}
where $\chi^{\alpha}_{i} (\alpha=1,2, \ i = 1,2,3)$ is indicated by the dots in Figs.~\ref{fig:Hallplot_all}(b)--\ref{fig:Hallplot_all}(d). The analytical forms are written as
\begin{align}
	\chi^{2}_{1}
	=&
	- \frac{C_{\zeta} x (x^{2} - 1)^{2}}{64 \gamma_{0}^{4} |x| (x^{2} + 1)^{4}}
	+ \mathcal{O} \left( \frac{1}{\gamma_{0}^{3}} \right),
	\label{eq:chi21}
	\\
	\chi^{2}_{2}
	=&
	\frac{3 \mathrm{sgn}(\Delta) C_{\zeta} x^{2} (x^{2}-1)}{32 \gamma_{0}^{4}|x| (x^{2}+1)^{4}}
	+ \mathcal{O} \left( \frac{1}{\gamma_{0}^{3}} \right),
	\label{eq:chi22}
	\\
	\chi^{2}_{3}
	=&
	\frac{5 \mathrm{sgn}(\Delta) C_{\zeta} x^{2} (x^{2}-1)^{2}}{32 \gamma_{0}^{4} |x| (x^{2}+1)^{5}}
	+ \mathcal{O} \left( \frac{1}{\gamma_{0}^{3}} \right),
	\label{eq:chi23}
\end{align}
and
\begin{align}
	\chi^{3}_{1}
	=&
	- \frac{3 \mathrm{sgn}(\Delta) C_{\zeta} x (x^{2}-1)^{2}}{128 \gamma_{0}^{4} |x| (x^{2}+1)^{4}}
	+ \mathcal{O} \left( \frac{1}{\gamma_{0}^{3}} \right),
	\label{eq:chi31}
	\\
	\chi^{3}_{2}
	=&
	- \frac{25 |x|(x^{2}-1)(3x^{4}-5x^{2}+3)}{512\gamma_{0}^{6}(x^{2}+1)^{6}}
	+ \mathcal{O} \left( \frac{1}{\gamma_{0}^{5}} \right),
	\label{eq:chi32}
	\\
	\chi^{3}_{3}
	=&
	- \frac{C_{\zeta}|x|(x^{2}-1)(21x^{4}+216x^{2}-205)}{4096\gamma_{0}^{6}(x^{2}+1)^{6}}
	+ \mathcal{O} \left( \frac{1}{\gamma_{0}^{5}} \right),
	\label{eq:chi33}
\end{align}
which are indicated by the lines in Figs.~\ref{fig:Hallplot_all}(b)--\ref{fig:Hallplot_all}(d).
Note that $\chi^{2}_{i}$ is an odd function of gap $\Delta$; this does not contradict the result of the symmetry consideration.

These results differ from those of ordinary ferromagnetic metals in that it is higher-order than the second order of relaxation time 
due to the perturbative treatment of the wavenumber of the magnetic textures. 
It is known that for ordinary ferromagnetic metals, the higher-order contribution is much smaller than 
that of the second-order contribution arising from the diffusion ladder vertices\cite{Nakazawa2019}.
On the other hand, the vertex corrections give only a quantitative contribution to the transport coefficients of Dirac fermions, 
which is comparable with the terms without the vertex corrections\cite{sakai14}. 
Therefore, our results, including higher-order terms of the relaxation time, 
would be comparable to the terms including the vertex corrections with the second-order relaxation time.

\begin{figure}[t]
	\centering
	\includegraphics[width=8.5cm]{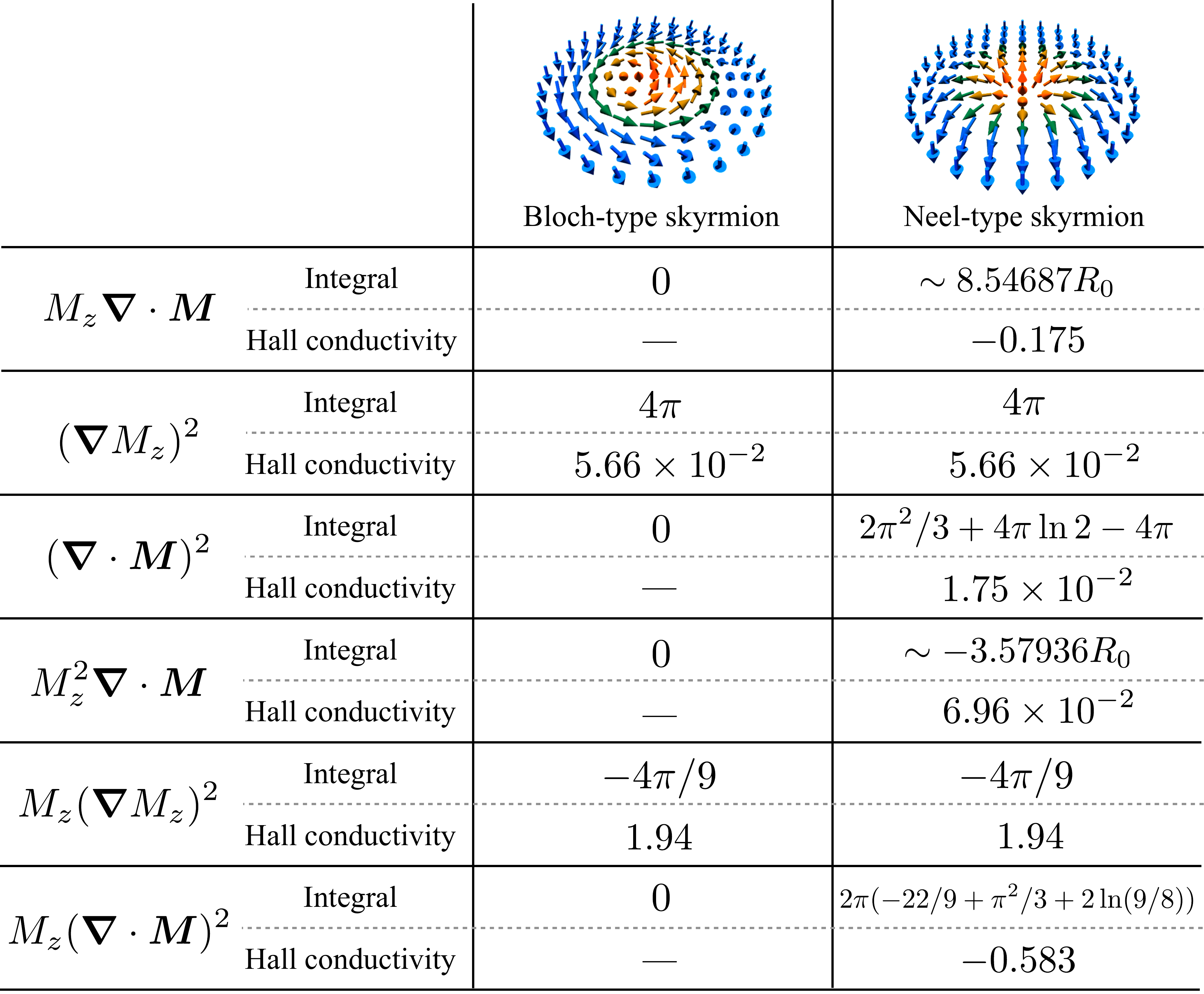}
	\caption{(Color online) Integrated values of magnetization and the Hall conductivities for Bloch-type and N\'{e}el-type skyrmions. 
		For each magnetization, the integrated value ($\int d \boldsymbol{r} M_z \boldsymbol{\nabla} \cdot \boldsymbol{M}$, etc.) is shown in the upper panel
		and the corresponding Hall conductivity is shown in the lower panel in units of $e^{2}/h$.
		We estimate Hall conductivities for $\mu/\Delta = 4$.
	}
	\label{fig:Hallplot_result_list}
\end{figure}

\textit{Estimation of Hall conductivity}.--
We estimate the Hall conductivity for a skyrmion lattice.
We note that the first order in magnetization in Eq.~(\ref{eq:sigma_1}) does not contribute to the Hall conductivity in the skyrmion lattice because it can be rewritten as the integral on the boundary of the system by performing partial integration. 
Then, we consider the second order and third order in magnetization in Eqs.~(\ref{eq:sigma_2}) and (\ref{eq:sigma_3}), respectively.
In addition, we assume two types of skyrmions, i.e., Bloch-type and N\'{e}el-type skyrmions, which are given by
\begin{align}
	\bm{M}_{\rm B}
	=
	\begin{pmatrix}
		- \sqrt{1 - M_{z}^{2}(\rho)} \sin \phi \\
		\sqrt{1 - M_{z}^{2}(\rho)} \cos \phi \\
		M_{z} (\rho)
	\end{pmatrix},
\
	\bm{M}_{\rm N}
	=
	\begin{pmatrix}
		\sqrt{1 - M_{z}^{2}(\rho)} \cos \phi \\
		\sqrt{1 - M_{z}^{2}(\rho)} \sin \phi \\
		M_{z} (\rho)
	\end{pmatrix},
	\label{eq:MB_MN}
\end{align}
where $(\rho, \phi)$ denotes real-space polar coordinates, and $\bm{M}_{\rm B}$ and $\bm{M}_{\rm N}$
correspond to Bloch-type and N\'{e}el-type skyrmions, respectively.
$M_{z}(\rho)$ is the $z$ component of magnetization, which depends only on $\rho$.
In addition, we assume that $M_{z}(\rho)$ has the Gaussian form,
$M_{z}(\rho) = 2 {\rm e}^{-\rho^{2}/R_{0}^{2}} - 1$, where $R_{0}$ corresponds to the radius of a skyrmion.
The space integrals of magnetization are shown in Fig.~\ref{fig:Hallplot_result_list}.

\begin{figure}[t]
	\centering
    \includegraphics[width=6cm]{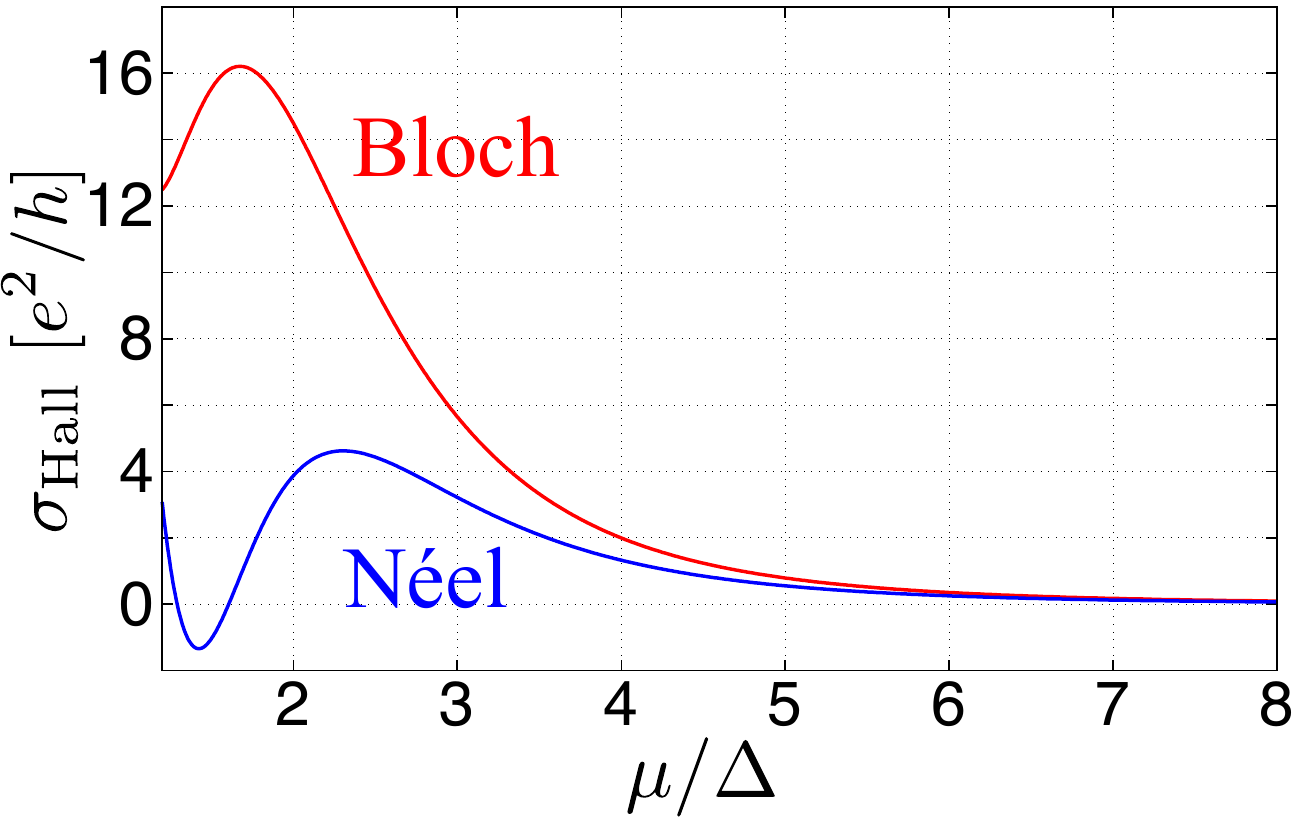}
	\caption{(Color online) Hall conductivity $\sigma_{\rm Hall}^{2} + \sigma_{\rm Hall}^{3}$
		for Bloch-type and N\'{e}el-type skyrmions with respect to chemical potential $\mu/\Delta$ in units of $e^{2}/h$.
	}
	\label{fig:Hallplot_totalHall}
\end{figure}

We use the following values to estimate the Hall conductivity: 
$J_{\perp} = J_{z} = 5$ meV, $\Delta = 8$ meV, $v_{\rm F} = 0.5 \times 10^6$ m/s,
and skyrmion density $n_{\rm s} = 1.0 \times 10^{7}$ cm$^{-2}$\cite{araki17}.
We use $n_{\rm s}$ to obtain the lattice spacing $a_{0} \simeq 1/(2\sqrt{n_{\rm s}}) \sim 3 \times10^{-4}$ cm,
and we assume $R_{0} = 300$ nm, which are reasonable comparing with experimental results \cite{Qin2018a}.
The chemical potential is fixed at $\mu/\Delta = 4$ and $\gamma_{0} = 0.05$.
Note that to ensure the justification of the approximation in our calculation, 
we should keep the length scales of skyrmion (skyrmion lattice spacing $a_{0}$ and radius $R_{0}$) 
are much larger than $\lambda_{\bm{\mathrm{F}}}=2 \pi /k_{\mathrm{F}}$ 
where $\lambda_{\mathrm{F}}$ and $k_{\mathrm{F}}$ are respectively Fermi wave length and Fermi wave number. 
For example, $\lambda_{\mathrm{F}} \simeq 67$ nm at $\mu/\Delta = 4$ is much smaller than $a_{0}$ and $R_0$.
In this condition, we obtain
$\chi^{2}_{1} \simeq - 6.76$, $\chi^{2}_{2} \simeq 10.7$, $\chi^{2}_{3} \simeq 15.6$,
$\chi^{3}_{1} \simeq -10.3$, $\chi^{3}_{2} \simeq -5382$, and $\chi^{3}_{3} \simeq -333$.
The estimated values of the Hall conductivity are also shown in Fig.~\ref{fig:Hallplot_result_list}.
$\bm{\nabla} \cdot \bm{M}$ is always zero for Bloch-type skyrmions, 
whereas it has a finite value for N\'{e}el-type skyrmions.
Figure \ref{fig:Hallplot_totalHall} shows the Hall conductivity, $\sigma_{\rm Hall}^{2} + \sigma_{\rm Hall}^{3}$, for Bloch-type and N\'{e}el-type skyrmions with respect to chemical potential $\mu / \Delta$.
The magnitude is of the order of $e^2/h$, which is considerably large compared to the normal and uniform-magnetization-induced Hall conductivities. 
The absolute value of the Hall conductivity for Bloch-type skyrmions is larger than that
for N\'{e}el-type skyrmions for a wide range of chemical potentials, whereas the sign of the Hall conductivity reverses at $\mu/\Delta < 2$ for N\'{e}el-type skyrmions.

\textit{Discussion}.--
We emphasize that the Hall conductivity in the current system depends on the type of skyrmions. 
The conventional topological Hall effect, which depends only on the skyrmion number \cite{conventional} $(1/4 \pi) \int {\rm d} \bm{r} \bm{M} \cdot (\partial_{x} \bm{M} \times \partial_{y} \bm{M})$ \cite{Ye1999, Bruno2004, Neubauer2009, Nakazawa2018}, is eliminated by the ``gauge'' symmetry caused by the linear dispersion of a Dirac fermion. 
The other terms that depend on the shape of the texture can be induced by spin--orbit coupling.

Recently, it has been found that there are correction terms to the THE written in the first order in the gradient of magnetic textures. 
The previous study\cite{Lux2020} shows that those are induced by an intrinsic mechanism, i.e., they do not depend on impurities. 
Our results include the first-order spatial derivative ($M_{z} \bm{\nabla} \cdot \bm{M}$ and $M_{z}^{2} \bm{\nabla} \cdot \bm{M}$)
depending on impurities, which implies that our results are different contributions from the results of the previous study. 
We also show the existence of not only the first-order but also higher-order contributions other than the form of skyrmion number.

Our result is not limited to skyrmions. It can be applied to general magnetic textures, including chiral domain walls \cite{Emori2013} and merons. 
Although it is difficult to experimentally observe the magnetic textures on the surface (or at the junction interface) of a topological insulator directly, our results suggest the possibility of obtaining information about magnetic textures by measuring the Hall conductivity, as the Hall conductivity varies with the type of skyrmions.

Furthermore, we can gain insight into the spin torque\cite{Slonczwski2005,sakai14,Mellnik2014} as 
the ``reaction'' of the topological Hall effect. 
The spin torque on the surface of a topological insulator is proportional to $\boldsymbol{\nabla} \cdot \bm{M}$ in first order in magnetization and spatical differentiation. 
In contrast, our results suggest that higher order in magnetization and differentiation, such as $\nabla^{2} M_{z}$, may also provide finite contributions in addition to $\boldsymbol{\nabla} \cdot \bm{M}$. 
As velocity operator $j_{\alpha}$ is written as
$j_{\alpha} = v_{\rm F} (\hat{z} \times \bm{s})_{\alpha}$ using the spin operator, 
coefficient $\chi_{\rm st}$ of term $\bm{t} = \chi_{\rm st} \hat{z} \times\bm{E}$, 
which is a part of the spin torque, is proportional to $\sigma_{\mathrm{Hall}}$ as
$\chi_{\rm st} = - (J_{\perp} M_{z}/v_{\rm F}) \sigma_{\rm Hall}$, where $\sigma_{\mathrm{Hall}}$ is a function of $\boldsymbol{\nabla}^n\boldsymbol{M}^m$, as derived above.  
This relation is proved in Appendix B.

In particular, the terms in $\sigma_{\rm Hall}$ that are proportional to $(\bm{\nabla} M_z)^2$ and $M_z (\bm{\nabla} M_z)^2$ are interesting because they depend only on the $z$ component of magnetization 
and provide a finite contribution, even when there is Ising-like coupling between magnetization and electrons. 
In other words, even for Ising-like interactions,
we can detect magnetic textures and control their dynamics using the magnetic-texture-induced Hall effect and spin torque, which can be applied to unconventional spintronic devices.
An interesting extension of our study is for topological superconductors, on which Majorana fermions live. 
The surface Majorana fermions on time-reversal-invariant topological superconductors have only Ising degrees of freedom for spins, 
called Majorana Ising spins. 
Furthermore, an octupole \cite{kobayashi19} and an (electric) quadrupole \cite{yamazak20} of Majorana fermions can be coupled to a magnetic texture. We expect that the magnetic-texture-induced Hall effect is used to detect Majorana fermions and their magnetic degrees of freedom; this will be discussed in a future work.


\textit{Conclusion}.--
We have theoretically investigated the Hall effect on topological-insulator surfaces due to magnetic textures.
From symmetry considerations, we have shown that the magnetic textures that contribute to the Hall conductivity are limited to 8 forms up to third order in magnetization and second order in spatial differentiation. 
In contrast, the conventional topological Hall effect, which is proportional to the skyrmion number, does not occur.
The coefficients of each term are calculated based on the linear response theory, and they provide a finite contribution to the Hall conductivity.
These results apply not only to conventional skyrmion systems but also to more general magnetic structures.
In addition, they are expected to be applied to unconventional spintronic devices.

\begin{acknowledgments}
    The authors are grateful to K. Nakazawa, J. J. Nakane, Y. Imai, S. Oyama and D. Nakamura for fruitful discussions.   
	A.Y. was supported by JSPS KAKENHI (Grants Nos.~JP20K03835 and JP20H04635) and the Sumitomo Foundation (190228). 
\end{acknowledgments}


\appendix

\section{Particle-hole symmetry for response functions}

Here we derive particle-hole symmetry for response functions. 
Denote the Hamiltonian as $H(\boldsymbol{X})$, where 
$\boldsymbol{X}$ is a set of parameters such as a magnetization with a texture, $\boldsymbol{X} = \boldsymbol{M}(\boldsymbol{x})$. 
The system is assumed to respect particle-hole symmetry as
\begin{align}
	C H(\boldsymbol{X}) C^{-1} = -H(\bar{\boldsymbol{X}}).
\end{align}
Let $\ket{\alpha; \boldsymbol{X}}$ be the state with the energy $E_\alpha(\boldsymbol{X})$.
The Schr\"{o}dinger equation is given by
\begin{align}
  H(\boldsymbol{X}) 
  \ket{\alpha; \boldsymbol{X}} = E_\alpha(\boldsymbol{X}) 
  \ket{\alpha; \boldsymbol{X}}.
\end{align}
The charge conjugation of the above equation is
\begin{align}
 H(\bar{\boldsymbol{X}})
 C
 \ket{\alpha; \boldsymbol{X}}
 = -E_\alpha(\boldsymbol{X}) 
 C \ket{\alpha; \boldsymbol{X}},
\end{align}
meaning that $H(\bar{\boldsymbol{X}})$ has the eigenvalue $-E_\alpha(\bar{\boldsymbol{X}})$ corresponding to $E_\alpha(\boldsymbol{X})$.
Then it is useful to set the following relation
\begin{align}
 &E_{\bar\alpha}(\bar{\boldsymbol{X}})
 := -E_\alpha(\boldsymbol{X}),
 \\
 &\ket{\bar\alpha; \bar{\boldsymbol{X}}}
 := C \ket{\alpha; \boldsymbol{X}},
 \\
 &H(\bar{\boldsymbol{X}})
 \ket{\bar\alpha; \bar{\boldsymbol{X}}}
 =
 E_{\bar\alpha}(\boldsymbol{X})
 \ket{\bar\alpha; \bar{\boldsymbol{X}}}.
 \label{eq:phs1}
\end{align}

Response function $K_{BA}(\omega; \mu, T, \boldsymbol{X})$ for physical quantity $B$ induced by an applied field $H_{\mathrm{ex}}(t) = -A F(t)$ has the form
\begin{align}
 & K_{BA}(\omega; \mu, T, \boldsymbol{X}) 
 \notag \\
 &= 
 \sum_{\alpha\beta}
 \frac{
 	e^{-\qty[E_\alpha(\boldsymbol{X}) -\mu N]/T} - e^{-\qty[E_{\beta}(\boldsymbol{X}) - \mu N]/T}
 }
 {Z(\mu, T, \boldsymbol{X})}
 \frac{B_{\alpha\beta}(\boldsymbol{X}) A_{\beta\alpha}(\boldsymbol{X})}{\omega + E_\alpha(\boldsymbol{X}) - E_\beta(\boldsymbol{X}) + i \gamma},
 \label{eq:K}
\end{align}
in the spectrum representation.
The matrix elements are defined by $A_{\alpha\beta}(\boldsymbol{X}) = \mel{\alpha; \boldsymbol{X}}{A}{\beta; \boldsymbol{X}}$ and satisfy particle-hole symmetry
\begin{align}
 &A_{\alpha\beta}(\boldsymbol{X})
 = 
 \mel{\bar\beta; \bar{\boldsymbol{X}}}{C A^\dag C^{-1}}{\bar\alpha; \bar{\boldsymbol{X}}}
 = 
 \eta_{C,A}
 A_{\bar\beta\bar\alpha}(\bar{\boldsymbol{X}}),
 \\
 &C A C^{-1} = \eta_{C,A} A,
 \
 \eta_{C,A} = \pm 1.
 \label{eq:phs_A}
\end{align}
The partition function also satisfies
\begin{align}
 Z(\mu, T, \boldsymbol{X})
 =& \sum_{\alpha}e^{-(E_\alpha(\boldsymbol{X})-\mu N)/T}
 = \sum_{\bar\alpha}e^{(E_{\bar\alpha}(\bar{\boldsymbol{X}})+\mu N)/T}
 \notag \\
 =& Z(-\mu, -T, \bar{\boldsymbol{X}}).
 \label{eq:phs_Z}
\end{align}
Substituting Eqs.~(\ref{eq:phs1}), (\ref{eq:phs_A}), and (\ref{eq:phs_Z}) into Eq.~(\ref{eq:K}), we obtain the particle-hole symmetry for the response function as
\begin{align}
 K_{BA}(\omega; \mu, T, \boldsymbol{X})
 =
 \eta_{C, A} \eta_{C, B}
 K_{BA}(\omega; -\mu, -T, \bar{\boldsymbol{X}}).
\end{align}

\section{Derivation of the relation between conductivity and spin torque}

The spin torque is written as
\begin{align}
	\bm{t} 
	= 
	J_{\perp} \bm{M} \times \langle \bm{s}^{\perp} \rangle
	+ J_{z} \bm{M} \times \hat{z} \langle s^{z} \rangle,
	\label{eq:sd_torque}
\end{align}
where $\bm{M}$ is the localized spin, $\bm{s}$ is the spin operator of conduction electrons 
and $\bm{s}^{\perp} = \bm{s} - (\hat{z} \cdot \bm{s}) \hat{z}$.
We define the correlator $\langle A;B \rangle$ for two arbitary operators $A,B$ as
\begin{align}
	& \langle A ; B \rangle
	=
	\lim_{\omega \to 0} \frac{K_{AB}(\omega + i 0) - K_{AB}(0)}{i \omega},
	\label{eq:correlator}
	\\
	& K_{AB} (i \omega_{\lambda})
	=
	\int_{0}^{1/T} \mathrm{d} \tau \mathrm{e}^{i \omega_{\lambda} \tau}
	\langle T_{\tau} A(\tau) B \rangle.
	\label{eq:KAB}
\end{align}
By using this correlator we can write the expect values of the spin and the current within the linear response of electric field as
\begin{align}
	\langle j_{\alpha} \rangle = \langle j_{\alpha} ; j_{\beta} \rangle E_{\beta},
	\quad
	\langle s^{\alpha} \rangle = \langle s^{\alpha} ; j_{\beta} \rangle E_{\beta}.
\end{align}
This leads the relation $\sigma_{\alpha \beta} = \langle j_{\alpha}; j_{\beta} \rangle$.
The current operator is written by using the spin operator as $\bm{j} = v_{\rm F} (\hat{z} \times \bm{s})$,
we obtain the relation
\begin{align}
	\langle s^{x} \rangle
	=&
	\frac{1}{v_{\rm F}} \langle j_{y} \rangle
	=
	\frac{1}{v_{\rm F}} \langle j_{y} ; j_{x} \rangle E_{x} + \frac{1}{v_{\rm F}} \langle j_{y}; j_{y} \rangle E_{y}
    \notag \\
	=&
	\frac{\sigma_{yx}}{v_{\rm F}} E_{x} + \frac{\sigma_{yy}}{v_{\rm F}} E_{y},
	\label{eq:sx}
	\\
	\langle s^{y} \rangle
	=&
	- \frac{1}{v_{\rm F}} \langle j_{x} \rangle
	=
	- \frac{1}{v_{\rm F}} \langle j_{x} ; j_{x} \rangle E_{x} - \frac{1}{v_{\rm F}} \langle j_{x}; j_{y} \rangle E_{y}
    \notag \\
	=&
	- \frac{\sigma_{xx}}{v_{\rm F}} E_{x} - \frac{\sigma_{xy}}{v_{\rm F}} E_{y}.
	\label{eq:sy}
\end{align}
Here we introduce some symbols for simplicity as
\begin{align}
	& \frac{1}{2} \left( \sigma_{xy} + \sigma_{yx} \right) = \sigma_{\rm AMR},
	\label{eq:sigma_AMR}
	\\
	& \frac{1}{2} \left( \sigma_{xy} - \sigma_{yx} \right) = \sigma_{\rm Hall},
	\label{eq:sigma_Hall}
	\\
	& \frac{1}{2} \left( \sigma_{xx} + \sigma_{yy} \right) = {\sigma}_{\ell},
	\label{eq:sigma_ell}
	\\
	& \frac{1}{2} \left( \sigma_{xx} - \sigma_{yy} \right) = \sigma_{d},
	\label{eq:sigma_d}
\end{align}
then we can rewrite Eq. (\ref{eq:sd_torque}) by substituting (\ref{eq:sx}), (\ref{eq:sy}) 
and (\ref{eq:sigma_AMR}) - (\ref{eq:sigma_d}) as
\begin{align}
	\bm{t}^{\perp}
	= &
	\frac{J_{\perp} M_{z}}{v_{\rm F}} 
	\left[ 
		- \sigma_{\rm Hall} (\hat{z} \times \bm{E})
		+ \sigma_{\rm AMR} (\hat{z} \times \bar{\bm{E}})
		+ {\sigma}_{\ell} \bm{E}
		+ {\sigma}_{d} \bar{\bm{E}}
	\right]
    \notag \\
	& - J_{z} \{ \langle s^{z} ; \bm{j} \rangle \cdot \bm{E} \} (\hat{z} \times \bm{M}),
	\label{eq:t_perp}
	\\
	t_{z}
	= & 
	- \frac{J_{\perp}}{v_{\rm F}}
	\left[
		\sigma_{\rm Hall} (\bm{M} \times \bm{E})_{z}
		- \sigma_{\rm AMR} (\bm{M} \times \bar{\bm{E}})_{z}
    \right.
    \notag \\
    & 
    \left.
		+ {\sigma}_{\ell} \bm{M} \cdot \bm{E} + \sigma_{d} \bm{M} \cdot \bar{\bm{E}}
	\right],
	\label{eq:t_z}
\end{align}
where $\bm{t}^{\perp} = \bm{t} - (\hat{z} \cdot \bm{t}) \hat{z}$ and  $\bar{\bm{E}} = (E_{x}, - E_{y})$.

\bibliography{17769}

\begin{thebibliography}{30}
\bibitem [{con()}]{conventional}%
  \BibitemOpen
  \href@noop {} {}\bibinfo {note} {The Hall conductivity without spin-orbit
  coupling must belong to the $A_2$ irrep of $C_{\infty v}$ in the real space
  and the $A_1$ irrep in the spin space. Only the skyrmion number meets this
  condition.}\BibitemShut {Stop}%
\end{thebibliography}
\bibliographystyle{apsrev4-1}

\clearpage

\end{document}